\documentclass[prl,showpacs]{revtex4}

\usepackage{epsfig}
\usepackage{color}

\usepackage{graphicx}

\usepackage{amssymb,amsfonts,amsmath}
\usepackage{amssymb,amssymb,graphics,graphicx,amsfonts,amsmath,empheq}
\usepackage{pdfpages}

\def\e{\mathrm{e}}
\begin{document}

\title{Kinetic vs. energetic discrimination in biological copying}

\author{Pablo Sartori$^1$ and Simone Pigolotti$^2$.}
\affiliation{$^1$Max Planck Institute for the Physics of Complex
  Systems. Noethnitzer Strasse 38 , 01187, Dresden,
  Germany. $^2$Dept. de Fisica i Eng. Nuclear, Universitat Politecnica
  de Catalunya Edif. GAIA, Rambla Sant Nebridi s/n, 08222 Terrassa,
  Barcelona, Spain.}

\begin{abstract}
  We study stochastic copying schemes in which discrimination between
  a right and a wrong match is achieved via different kinetic
  barriers or different binding energies of the two matches. We
  demonstrate that, in single-step reactions, the two discrimination
  mechanisms are strictly alternative and can not be mixed to further
  reduce the error fraction.  Close to the lowest error limit, kinetic
  discrimination results in a diverging copying velocity and
  dissipation per copied bit. On the opposite, energetic
  discrimination reaches its lowest error limit in an adiabatic regime
  where dissipation and velocity vanish.  By analyzing
  experimentally measured kinetic rates of two DNA polymerases, T7 and
  Pol$\gamma$, we argue that one of them operates in the kinetic and
  the other in the energetic regime. Finally, we show how the two
  mechanisms can be combined in copying schemes implementing error
  correction through a proofreading pathway.
\end{abstract}

\maketitle

Living organisms need to process signals in a fast and reliable
way. Copying information is a task of particular relevance, as it is
required for the replication of the genetic code, the transcription of
DNA into mRNA, and its translation into a protein. Reliability is
fundamental, since errors can result in the costly (or harmful)
production of a non-functional protein. Indeed, cells have
developed mechanisms to reduce the copying error rate $\eta$
to values as low as $\eta\sim10^{-4}$ for protein
transcription-translation \cite{zaher09} and $\eta\sim10^{-10}$ for
DNA replication \cite{johnson93}. Such mechanisms include multiple
discrimination steps \cite{zaher09,johnson93} and pathways to undo
wrong copies as in proofreading \cite{hopfield74,ninio75,freter,leibler} or
backtracking \cite{shaevitz03}.

Biological information is copied by thermodynamic machines that
operate at a finite temperature. There is agreement that this fact
alone implies a lower limit on the error rate. However, contrasting
results have been obtained regarding the nature of this limit. In
particular, it is not clear when it is reached in a slow
and quasi-adiabiatic regime, or in a fast and dissipative one.  As
clarified by Bennett \cite{bennett82}, information can be copied
adiabatically. Indeed, the copying scheme proposed in Hopfield's
seminal proofreading paper \cite{hopfield74} reaches its minimum error
at zero velocity and zero dissipation
\cite{ehrenberg08}. In contrast, a copolymerization model proposed few
years later by Bennett \cite{bennett79,andrieux08,vandenbroeck10},
achieves its minimum error in a highly dissipative regime, where velocity and 
dissipation diverge.  Some of the biological
literature has favoured that the minimum error is
achieved in near-equilibrium conditions \cite{ehrenberg08}.  This view
is however not unanimous \cite{thompson82}. Recent biophysical
literature supports a dissipative minimum error limit
\cite{bennett08,andrieux08,jarzynski08,vandenbroeck10}.  Similar
disagreements are also present in models including proofreading.
The proofreading model in \cite{bennett79} dissipates
systematically less than the corresponding copying, while in
other models \cite{hopfield74,ninio75}, at low errors,
dissipation comes mainly from the proofreading step.

\begin{figure}[htb]
\centerline{\includegraphics[width=\textwidth]{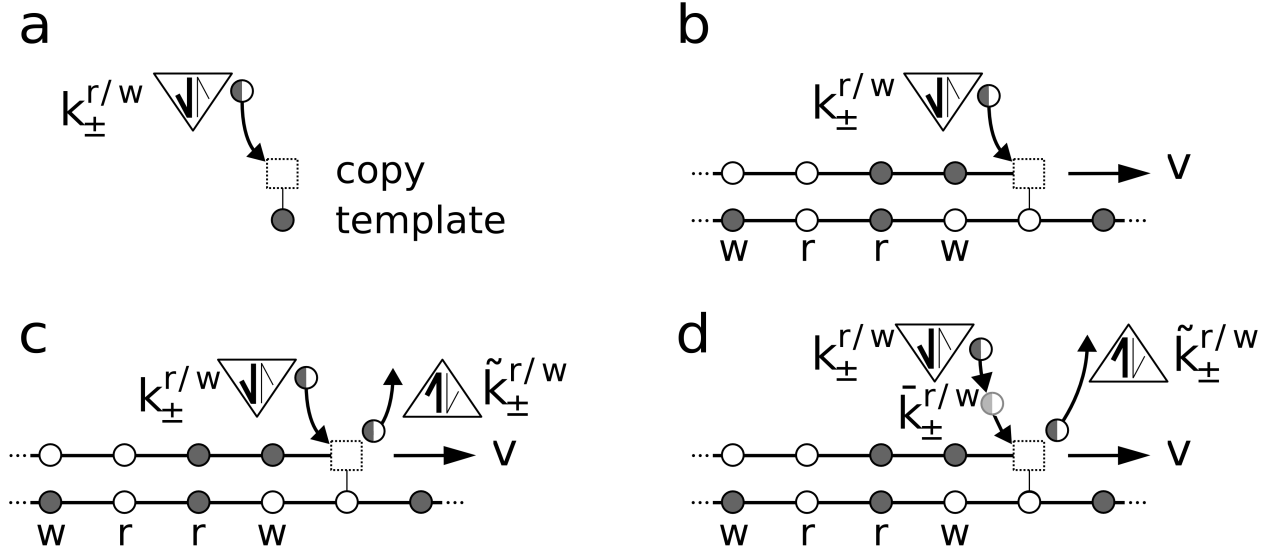}}
\caption{a) Copying of one bit. The lower-vertex triangle represents a
  bio-machine, such as a polymerase, binding or unbinding right and
  wrong matches with rates $k_{\pm}^{r/w}$. b) Copolymerization. A
  template strand (bottom) is copied into a new strand (top). Right
  (r) and wrong (w) matching monomers are added and removed 
  ($\pm$) with rates
  $k_{\pm}^{r/w}$ resulting in a growth velocity $v$.  c) \& d)
  Proofreading schemes. The polymerase is assisted by an exonuclease which 
  removes wrong matches,
  represented by an upper-vertex triangle and characterized by
  $\tilde{k}_{\pm}^{r/w}$. In d), copies are made via an intermediate
  state characterised by $\bar{k}_{\pm}^{r/w}$.
\label{schemes}}
\end{figure}

In this Letter, we show how these contrasting results can be
rationalized noting that a copy can be performed either
discriminating through binding energies adiabatically, {\em energetic
  discrimination}, or discriminating through binding barriers
dissipatively, {\em kinetic discrimination}.  We begin by presenting a
model for copying a single bit of information in the spirit of those
proposed in \cite{bennett82,parrondo,sagawa,granger}, see Fig.
\ref{schemes}a. A bio-machine such as a polymerase binds and unbinds
monomers of different species to a template, trying to match it.  We
then move to the case of copolymerization, Fig.  \ref{schemes}b,
where a polymerase assembles a polymer chain to match a template
strand. Finally, we discuss two proofreading schemes,
Figs. \ref{schemes}c and \ref{schemes}d, where the polymerase is
assisted by an exonuclease that tends to remove wrong matches.

\begin{figure}[htb]
\centerline{\includegraphics[width=\textwidth]{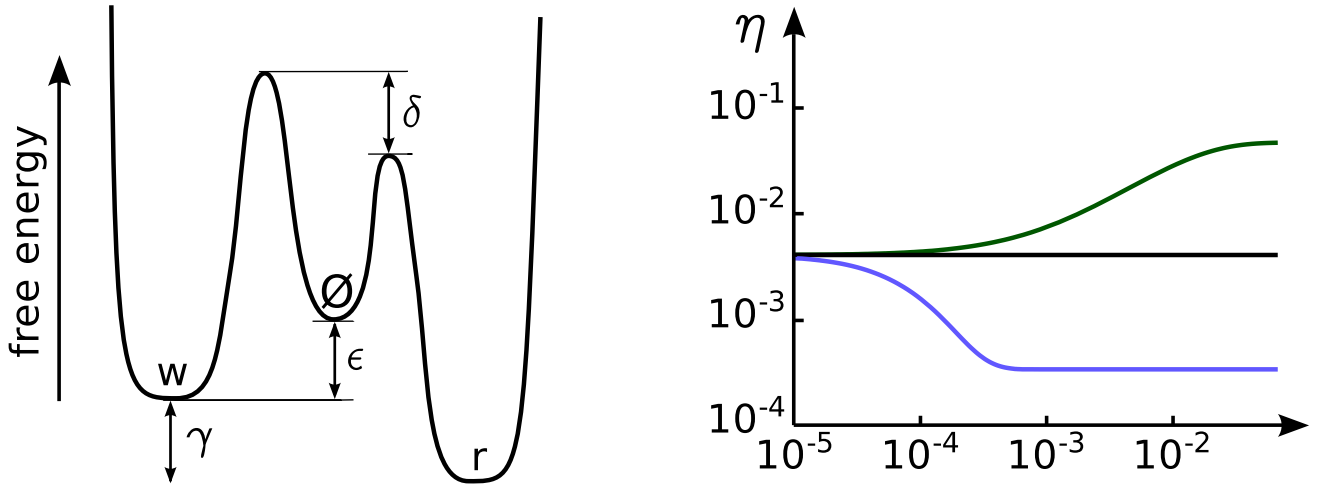}}
\caption{a) Energy diagram for copying rates.  The barrier height
  difference $\delta$ biases right additions, while the energy difference 
  $\gamma$  favours wrong removals. The right and wrong chemicals drivings 
  are $\epsilon$ and$\epsilon-\gamma$.  b)
  Time-evolution of the error in the single bit copy for three parameter
  choices: $\gamma<\delta$ (green curve, $\gamma=3$ and $\delta=5.5$),
  $\gamma>\delta$ (blue curve, $\gamma=8$ and $\delta=5.5$), and
  $\gamma=\delta=5.5$ (black curve). The other parameters are
  $\epsilon=5$ and $\omega=4$.
\label{errtime}}
\end{figure}

{\em Stochastic copying strategies of a single bit.}  The copying
machine is described as a three-states system. Two are bound states in
which the right ($r$) or wrong ($w$) molecule is attached to the
machine. The third is a ``blank'' state ($\emptyset$), representing
the unbound state of the machine before a matching is done.  To help physical
intuition and following \cite{bennett79}, we define the rates from the
free energy landscape in Fig. \ref{errtime}a.  Right and wrong
matching are characterized by a difference in barrier height $\delta$,
and in the energy of the final states $\gamma$. The energy $\epsilon$
is a chemical driving.  All energies are in units of $k_BT$,
  where $k_B$ is the Boltzmann constant and $T$ the temperature.

The four rates $k_{\pm}^{r/w}$ connecting the unbound state with the right
and wrong states can be written as Kramers rates from energy
  barriers of Fig. \ref{errtime}a as:
\begin{equation}
k_+^r=\omega \e^{\epsilon+\delta} \;\; ; \;\;k_-^r=\omega \e^\delta
\;\; ; \;\; k_+^w=\omega \e^\epsilon \;\; ; \;\; k_-^w=\omega 
\e^\gamma .\label{ratescop}
\end{equation}
where $\omega$ is an overall rate scale.
The master
equation for the probabilities $p_r$ and $p_w$
of finding the system in the right or wrong state  reads
\begin{eqnarray}\label{eq_model1}
\dot{p}_r &=& (1-p_r-p_w)k^r_+ - k_-^rp_r\\\nonumber 
\dot{p}_w&=&(1-p_r-p_w)k^w_+ - k_-^wp_w 
\end{eqnarray} 
where $p_\emptyset$ has been eliminated by normalization.  We study
the time-dependent error rate $\eta(t)=p_w(t)/[p_r(t)+p_w(t)]$ for the
system prepared in the unbound state, $p_r(t=0)=p_w(t=0)=0$.  At short
times, $t\ll\omega^{-1}$, one has $p_r \approx tk_+^r$ and $p_w
\approx tk_+^w$. To shorten notation, we define the function
$f(x)=e^{-x}/(1+e^{-x})$ mapping energies into errors.  The short-time
error is then $\eta(t \to 0)=f(\delta)$.  In the opposite limit of
$t\gg \omega^{-1}$, the system reaches equilibrium so that $\eta(t \to
\infty)=f(\gamma)$ by detailed balance.  At intermediate times, one
can demonstrate from the analytical solution of Eqs. \ref{eq_model1}
that $\eta(t)$ is a monotonic function for any choice of rates (see
\cite{supplements}): increasing with time when $\delta>\gamma$
(i.e. $f(\delta)<f(\gamma)$), and decreasing when $\gamma>\delta$
(i.e. $f(\delta)>f(\gamma)$).  For $\delta = \gamma$, the error is
time-independent.  The three cases are shown in Fig. \ref{errtime}b.

To maximize accuracy, the copying reaction must be arrested when
$\eta(t)$ is at its minimum value, quenching the system into either a
right or wrong copy outcome. In an enzymatic reaction, this
corresponds to the irreversible transformation of bound states into
products \cite{hopfield74}. In \cite{bennett82}, where bits are
encoded in ferromagnets, it corresponds to decoupling from an external
transverse field.  We define the {\em kinetic discrimination} regime
$\delta>\gamma$, where optimal accuracy requires stopping the process
as fast as possible. If $\gamma>\delta$, {\em energetic
  discrimination} regime, optimal accuracy is reached at very long
time, when the reaction reaches equilibrium.  In all cases, accuracy
can not be improved by combining the two mechanisms, as the lower
limit on the error is determined by either $\gamma$ or
$\delta$. Notice that in an energetic discrimination scheme, the
quench can be performed slowly, at no dissipation \cite{bennett82}. In
a kinetic scheme, the quench has to be fast and dissipative.

{\em Kinetic and energetic discrimination in copolymerization.}  In
copolymerization, a polymerase stochastically adds and removes
monomers to a tip of the growing copy strand, trying to match them
with those on the template strand (see Fig. \ref{schemes}b and
\cite{bennett82,andrieux08,vandenbroeck10,supplements}).  The model is
defined by the incorporation and removal rates of right $k_\pm^r$ and
wrong $k_\pm^w$ matching monomers, defined by Eq. (\ref{ratescop}) and
Fig. \ref{errtime}a.  The chemical drivings of the polymerase for
right and wrong bases are $\epsilon$ and $\epsilon-\gamma$. These bias
monomer addition over removal and ensure growth of the copied strand
at an average velocity $v\ge0$.  Monomer addition/removal and
polymerase forth/back stepping are thus tightly coupled  (relaxing this
has no effect on our results \cite{supplements}).  Previous studies on
copolymerization assumed iso-energetic strands, i.e.  $\gamma=0$
\cite{bennett79,bennett08,andrieux08,vandenbroeck10}.  We relax this
assumption and study how the copying velocity $v$, and the rate of
entropy production or dissipated chemical work $\dot{S}$ \cite{callen}
depend on the error rate $\eta$ for a general choice of $\delta$ and
$\gamma$.  It is straightforward to show that $v
=k_+^r-(1-\eta)k_-^r+k_+^w-\eta k_-^w $
\cite{bennett79,vandenbroeck10}, and also that $\dot{S}$ is given by
\begin{equation}
\dot{S} = v\Delta S= v (1-\eta)\epsilon +v \eta
  (\epsilon - \gamma) + v H(\eta) \label{epr} 
\end{equation} 
where $\Delta S=\dot{S}/v$ is the dissipation per added monomer, and
$H(\eta)=-\eta\log(\eta)-(1-\eta)\log(1-\eta)$ is the Shannon entropy
of the error rate $\eta$.  The first two terms in Eq. (\ref{epr})
represent the distinct chemical driving forces of right and wrong
bases, multiplied by the flux of right and wrong incorporated bases.
The last term of Eq. (\ref{epr}) corresponds to the information
entropy increase due to incorporation of errors, hence information, into
the chain \cite{bennett79,andrieux08,vandenbroeck10}.

\begin{figure}[htb]
\begin{center}
\includegraphics[width=\textwidth]{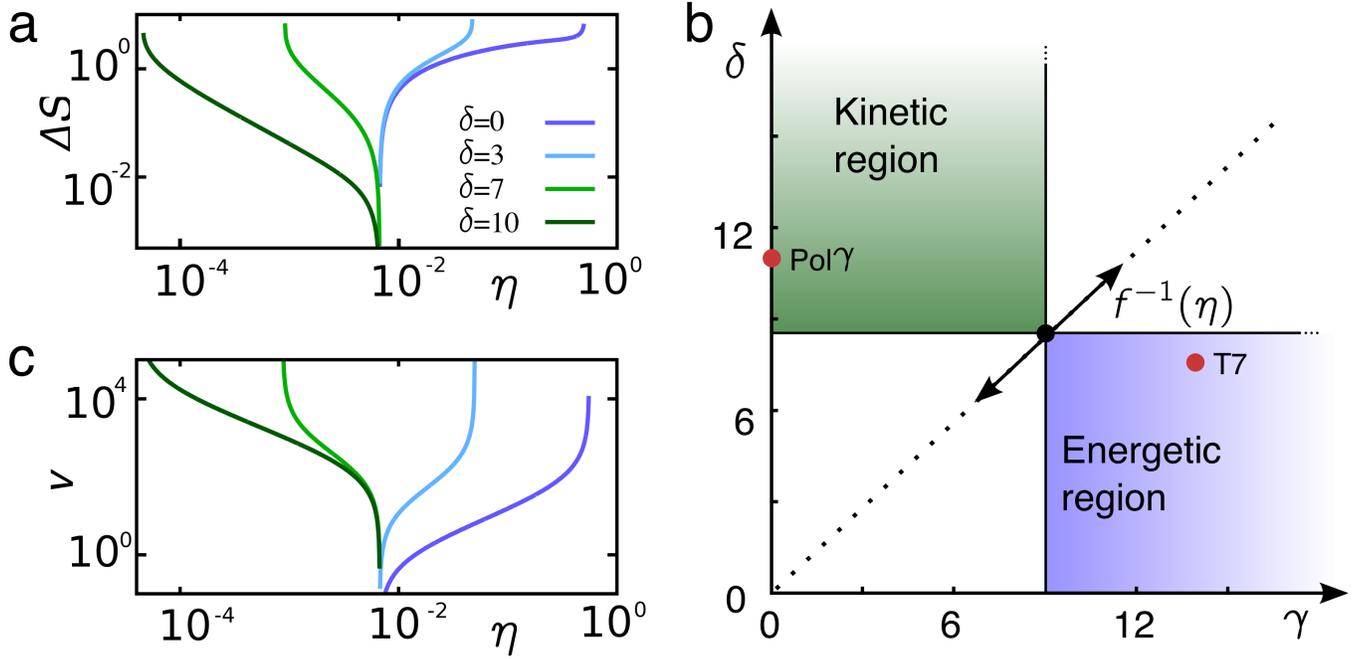}
\caption{a) Dissipation per step in copolymerization. In all curves
  $\gamma=5$, while $\delta$ varies. All curves tend to zero at
  $\eta=f(\gamma)\approx 6.7\cdot 10^{-3}$. Blue curves are in the
  energetic region, $\gamma>\delta$, while green curves are in the
  kinetic region $\delta>\gamma$.  b) $\gamma$-$\delta$ phase diagram,
  showing the kinetic and energetic discrimination regions compatible
  with an error $\eta\sim f(7)\approx 9\cdot 10^{-4}$, and estimated
  values of $(\gamma,\delta)$ for Pol$\gamma$ and T7. Tuning $\eta$
  shifts the limit $f^{-1}(\eta)$ of phase regions along the line
  $\gamma=\delta$. c) Behavior of the velocity $v$ for the
  parameter choices in a).
\label{copol_fig}}
\end{center}
\end{figure}

By imposing steady state flux conservation, we express $\epsilon$ in
terms of $(\eta,\delta,\gamma)$. Substituting, we obtain $\Delta
S(\eta,\delta,\gamma)$ and $v(\eta,\delta,\gamma)$ \cite{supplements},
presented in Fig. \ref{copol_fig}a and \ref{copol_fig}c for a fixed
value of $\gamma$ and different values of $\delta$.  The physical
range of admissible errors depends on $\gamma$ and $\delta$
(see \cite{supplements}) as
\begin{equation}\label{inequal}
  \min\left[f(\delta),f(\gamma)\right] < \eta <
  \max\left[f(\delta),f(\gamma)\right], 
\end{equation}
with $f(x)$ as previously defined. We now study the dissipative limit,
$\eta\rightarrow f(\delta)$, and the adiabatic limit, $\eta
\rightarrow f(\gamma)$.

When $\eta\rightarrow f(\delta)$, the chemical driving
diverges as $\epsilon \sim-\log|\eta-f(\delta)|$ \cite{supplements}.
Substituting into Eq. (\ref{epr}) shows that also $\Delta S$ diverges (see
Fig. \ref{copol_fig}a) as
\begin{equation}
\Delta S\sim\eta\epsilon \sim\eta \log|\eta-f(\delta)| \quad \mathrm{ for }
\quad \eta\rightarrow f(\delta).
\end{equation}
Since $\epsilon \gg 1$, the information entropy
$H(\eta)$ in Eq. (\ref{epr}) is negligible, and dissipation is
dominated by the chemical terms. As an effect of the strong driving,
the velocity diverges as $|\eta-f(\delta)|^{-1}$, see
Fig. \ref{copol_fig}c.

When $\eta\rightarrow f(\gamma)$, both $v$ and $\Delta S$ tend to
zero, see Fig. \ref{copol_fig}a and \ref{copol_fig}c: all the chemical
energy is invested in copying the information, none being wasted. The
chemical driving is then
\begin{equation}
 \epsilon = \log(1-\eta)  = \log[1-f(\gamma)] < 0\quad 
\mathrm{ for }\quad \eta=f(\gamma).
\end{equation}
Note that $\epsilon$ is small and {\em negative}, to compensate the
small positive entropic driving caused by $H(\eta)$ in
Eq. (\ref{epr}).  

By inverting Eq. (\ref{inequal}), the values of $\gamma$ and $\delta$
compatible with a given error $\eta$ must satisfy either
$\gamma<f^{-1}(\eta)<\delta$ or $\delta<f^{-1}(\eta)<\gamma$, with
$f^{-1}(x)=\log(1+1/x)$ the inverse of $f(x)$.  This defines the two
disconnetted {\em kinetic discrimination} ($\delta>\gamma$), and {\em
  energetic discrimination} ($\gamma>\delta$) regions of the
$(\gamma,\delta)$ plane in Fig. \ref{copol_fig}b.

In the kinetic region, both $\Delta S$ and $v$ diverge in the minimum
error limit, so that accuracy comes at the cost of high
dissipation. In the energetic region, accurate copying comes at the
cost of the copying velocity, which goes to zero in the adiabatic
minimum error limit. This fundamental difference is at the core of the
discrepancies between enzymatic copying models \cite{hopfield74} that
assumed lack of forward discrimination, $\delta=0$ in our language
(see \cite{supplements} for mapping), and copolymerization studies
\cite{bennett79,andrieux08,vandenbroeck10} that assumed iso-energetic
strands, $\gamma=0$.  Our results show that it is impossible to
interpolate between the two, as they belong to two separate regions of
parameter space.

{\em Operating regimes of T7 and Pol$\gamma$ polymerases.} We now
analyze two specific biological copying systems: DNA replication of
the phage T7 \cite{johnson93,hrlee}, and replication of human DNA by
Pol$\gamma$ \cite{tsai06}.  A recent experimental study \cite{tsai06}
points at the strong and asymmetric backward rates as the leading
discriminatory mechanism in T7. We derived from \cite{tsai06} the
copolymerization rates by assuming equilibrium nucleotide binding with
dissociation constants $K_r=28\mu$M and $K_w=200\mu$M for right and
wrong base matching.  Considering nucleotide concentrations in a range
of $[dNTP]\sim 0.5 - 50\mu$M we obtain the binding states
$1/(1+K_{r/w}/[dNTP])$.  Multiplying them by the forward rates (360Hz
and 0.2Hz for right and wrong bases respectively) we obtain
$k_+^{r/w}$. The backward rates are $k_-^r\approx2$Hz and
$k_-^w\approx0.04$Hz \cite{tsai06}.  These values give an error range
$\eta\sim10^{-6}-10^{-4}$, in agreement with \cite{johnson93}. Usual
estimates of the error assume linear binding, approximation valid for
low $[dNTP]$ and yielding the lowest end of the error range. The
velocities are $v\sim 5-250$bps (bases per second), in agreement with
the saturation rate measured in \cite{tsai06}. By inverting
Eqs. (\ref{ratescop}), we can infer $\gamma\approx14$ and
$\delta\approx8$.  Since $\gamma>\delta$, we conclude that T7 operates
in the energetic regime (see Fig. \ref{copol_fig}b).

DNA duplication by Pol$\gamma$ was analyzed in \cite{andrieux08} with
a variant of the copolymerization model, where different monomer
species are characterised by different rates. Agreement with
experimental data in \cite{hrlee} was obtained assuming that the copy
be iso-energetic ($\gamma=0$). We simplify the analysis in
\cite{andrieux08} by averaging over the different monomer
species. Using the same driving $\epsilon\approx5$ determined for T7,
we obtain $\delta\approx11$ and a range of error rates
$\eta\sim10^{-5}-10^{-3}$. In the limit of low $[dNTP]$ it agrees with
the estimates in \cite{hrlee,andrieux08}.  As $\gamma=0$, Pol$\gamma$
lies in the kinetic discrimination region
(Fig. \ref{copol_fig}b).  While here as in \cite{andrieux08}
  $\gamma=0$ was assumed for simplicity, a non-zero value of $\gamma$
  but smaller than $\delta$ would not alter our main conclusion.

 The estimates of $\delta$ and $\gamma$ above indicate that, while
  the two polymerases achieve a similar error rate $\eta$, they
  operate in different regimes, implying different tradeoffs. In T7,
lowering $[dNTP]$ (effectively, the chemical driving) can reduce the
error $\eta$. This also reduces the dissipation $\Delta S$, at the
cost of a smaller speed $v$. This situation is similar to that of the
blue curves in Figs. \ref{copol_fig}a and \ref{copol_fig}c. In
Pol$\gamma$, a smaller error requires a stronger driving, hence
dissipation \cite{andrieux08}. This gives a higher polymerization
rate, as in the green curves of Figs. \ref{copol_fig}a and
\ref{copol_fig}c.

\begin{figure}[htb]
\begin{center}
\includegraphics[width=\textwidth]{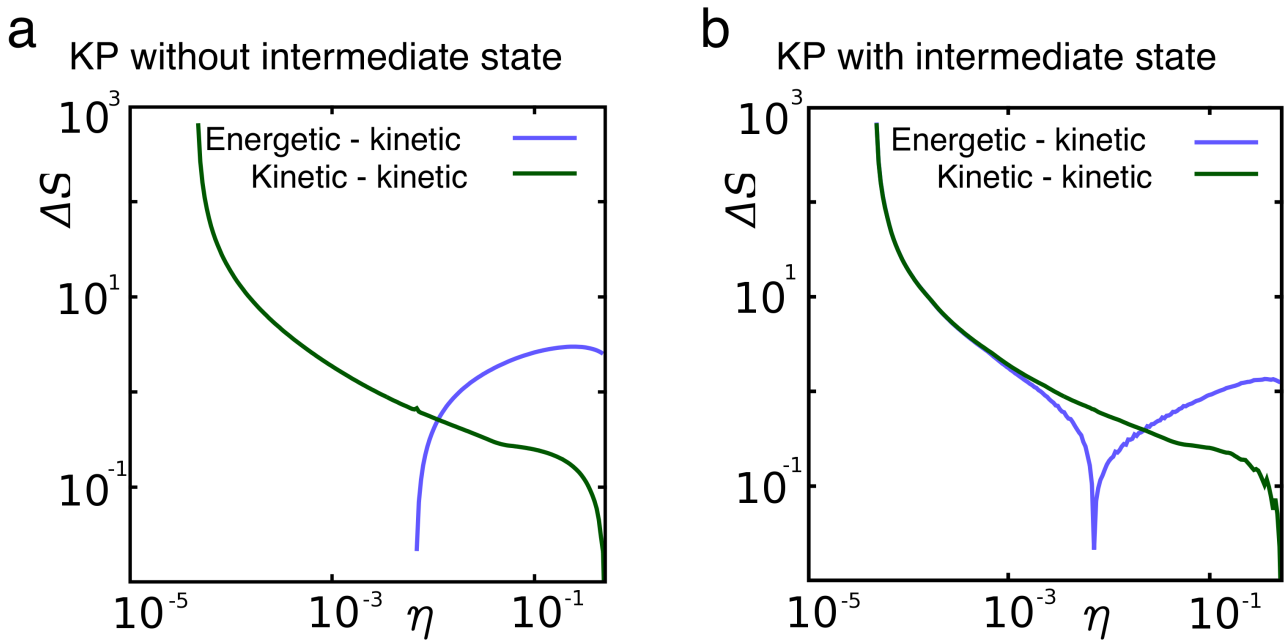}
\caption{a) Minimum dissipation in proofreading
  without an intermediate step, Fig. \ref{schemes}c. For both curves,
  $\gamma_p=0$ and $\delta_p=5$. In the case of energetic
  discrimination in copying and kinetic discrimination in
  proofreading (energetic-kinetic), the other parameters are
  $\gamma=5$, $\delta=0$. In the kinetic-kinetic case, we used
  $\gamma=0$ and $\delta=5$.  b) Minimum dissipation
   in  proofreading with an intermediate step, Fig. \ref{schemes}d.
   Parameters are as in a). In both
  panels, the expected minimum errors $f(10)$ and $f(5)$, depending on
  whether proofreading is effective or not, are marked.
  \label{proofread_fig}}
\end{center}
\end{figure}

{\em Combining copying strategies in proofreading schemes.} 
 We now explore the
possibility of combining the two mechanisms in multi-step copying
schemes involving a proofreading pathway.  In proofreading, an
initially copied base can be removed via an alternative pathway, see
Fig. \ref{schemes}c and \ref{schemes}d. Such erasing pathway is
characterized by a discrimination which, a priori, can be energetic
$\gamma_p$ or kinetic $\delta_p$, a distinct time-scale $1/\omega_p$,
and a (backward) driving $\epsilon_p$. In an effective proofreading
scheme, the minimal copying error of Eq. (\ref{inequal}) is reduced by
an additional proofreading factor, in principle energetic
$f(\gamma_p)$ or kinetic $f(\delta_p)$. We discuss two proofreading
schemes. In both of them, the proofreading rates
$\tilde{k}_{\pm}^{r/w}$ have the same structure as the copying ones,
apart from a backward driving \cite{supplements}. In the first,
Fig. \ref{schemes}c, the copying step is identical to that in the
copolymerization model, as in Bennett's proofreading model
\cite{bennett79}.  In the second, Fig.  \ref{schemes}d, the copying
step leads to an intermediate state, taken to its final form via rates
$\bar{k}_{\pm}^{r/w}$ without further discrimination, as in Hopfield's
model \cite{hopfield74}.  By imposing flux balance at the steady state
we solved both models analytically \cite{supplements}. We fixed the
discrimination factors, and for each error $\eta$ minimized $\Delta S$
over the remaining free parameters \cite{supplements}, obtaining the
curves of minimum dissipation vs. error in Fig. \ref{proofread_fig}.

As shown in \cite{supplements}, there are no regimes in any of the two
proofreading schemes where the error is lowered by the energetic
factor $f(\gamma_p)$, while error reduction by a kinetic proofreading
factor $f(\delta_p)$ is feasible, see
  Fig. \ref{proofread_fig}a and \ref{proofread_fig}b. Proofreading is
thus only effective when it operates in the kinetic regime.  This
result is consistent with Landauer's principle \cite{landauer61}, as
erasure of information (errors) constitutes an intrinsically
dissipative process.  Further, by looking at the minimum errors in
Fig. \ref{proofread_fig}, one can conclude that, while kinetic
proofreading is always effective when combined with kinetic copying
(green curves), it is only compatible with adiabatic copying when an
intermediate state is present (blue curves). This is a key difference
between the proofreading schemes in \cite{bennett79} and
\cite{hopfield74}: without an intermediate state it is impossible to
find a regime where copies are produced adiabatically, and undone very
quickly.  The combination of kinetic proofreading with adiabatic
copying step has the advantage of a lower dissipation (see
Fig. \ref{proofread_fig}b, green vs. blue lines).

In this Letter, we have shown how {\em each} copying step in
stochastic copying can be unambiguously classified into one of two
radically different classes, kinetic and energetic
discrimination.  These regimes are reminiscent of kinetic and
  thermodynamic control in chemistry, where however the two
  discrimination factors appear in parallel competing pathways \cite{organchem}. The
existence of an energetic regime in the copolymerization model
complements the view in the literature
\cite{bennett79,andrieux08,vandenbroeck10} that low copy errors are
achieved only in a highly dissipative regime. It also demonstrates how
entropy-driven growth, a phenomenon studied in the large error regime
\cite{bennett79,bennett08,andrieux08, jarzynski08}, can be exploited
to reliably copy information.  Copolymerization is thus compatible
with the principle of reversible computing stating that a copy can be
performed adiabatically \cite{bennett82}.  The analysis of two DNA
polymerases, T7 and Pol$\gamma$, shows that the first operates in the
energetic regime, while the second in the kinetic one. Both mechanisms
are thus used by biological systems. Finally, our study of
proofreading proves that the two regimes discussed here can be combined
in more complex copying schemes.

Our conceptual framework can be applied to a wider range of problems
related to stochastic discrimination. Examples are detection of
antigens by T-cell receptors \cite{mckeithan95}, and discrimination of
a binary input in neural dynamics \cite{wang02}.  At the sub-cellular
level, thermal fluctuations dominate and impose constraints on
biological tasks.  While the thermodynamics of bio-mechanical systems
such as molecular motors is well understood \cite{frank}, the role of
fluctuations in biological information processing such as bacterial
chemotaxis presents still many open questions \cite{yuhai}.  Our work
shows that the emerging trade-offs may be complex, and depend on the
region in parameter space where the system operates.

\begin{acknowledgments}
  This work was partially supported by a Max Planck Society
  scholarship (to P.S.) and a Ramon y Cajal Grant (to S.P.). We are
  grateful to A. Bernacchia, J. Garcia-Ojalvo, N. Mitarai, L. Granger,
  M. A. Mu\~noz and Y. Tu for a critical reading of the manuscript.
\end{acknowledgments}

\newpage

\section{Supplementary Information}

This document contains additional details and derivation of the
results presented in the Letter ``Energetic vs. kinetic discrimination
in biological copying''. The document is organized as follows.
Section 1 presents a full solution of the single bit copying model
(model A in Fig. 1 of the main text), and a demonstration that the
error is always a monotonic function of time. Section 2 details
results on the co-polymerization model (model B in the main
text). Section 3 illustrates the mapping between Hopfield's and
Bennett's copying schemes. Sections 4 and 5 present details on
the proofreading models (models C and D in the main text,
respectively). Finally, section 6 discusses copying when polymerase stepping and 
base copying are independent (mathematically a particular case of model D). 

\section{Stochastic copying of a single bit}

We wish to demonstrate that, for any choice of the rates, the solution
of the system of differential equations:

\begin{eqnarray}\label{threestate_sys}
\dot{p}_r&=&k^r_+(1-p_r-p_w)-k^r_- p_r\nonumber\\
\dot{p}_w&=&k^w_+(1-p_r-p_w)-k^w_- p_w.
\end{eqnarray}

with initial condition $p_r(t=0)=p_w(t=0)=0$ leads to a time dependent error
\begin{equation}
\eta(t)=\frac{p_w(t)}{p_r(t)+p_w(t)}.
\end{equation}
being a monotone function of time for all $t>0$. In particular,
$\eta(t)$ either strictly increasing, strictly decreasing or constant
depending on the choice of the parameters.

The solution of the system of equations (\ref{threestate_sys}) can be
obtained with standard methods. First of all, the steady state
solution is:
\begin{eqnarray}
p_{r,eq}&=&\frac{1}{1+\frac{k^r_-}{k^r_+}+\frac{k^r_-k^w_+}{k^r_+k^w_-}}\nonumber\\
p_{w,eq}&=&\frac{1}{1+\frac{k^w_-}{k^w_+}+\frac{k^w_-k^r_+}{k^w_+k^r_-}}.
\end{eqnarray}
Upon defining $\delta p_r=p_r-p_{r,eq}$ and $\delta p_w=p_w-p_{w,eq}$ the time-dependent
distances from the steady state, a lengthy but straightforward
calculation leads to
\begin{eqnarray}
\delta p_r(t)&=&\frac{N_-
\left\{-p_{r,eq}+\frac{p_{w,eq}}{2k^w_+}
  \left[q+\sqrt{q^2+4k^r_+ k^w_+}\right]\right\}e^{\lambda_+ t}}{4+\frac{q^2}{k^r_+ k^w_+}}\\
&+&\frac{N_+\left\{-p_{r,eq}+\frac{p_{w,eq}}{2k^w_+}
\left[q-\sqrt{q^2+4k^r_+ k^w_+}\right]\right\}e^{\lambda_- t}}{4+\frac{q^2}{k^r_+ k^w_+}}
\nonumber\\
\delta p_w(t)&=&\frac{-N_-\left(q+\sqrt{q^2+4k^r_+k^w_+}\right)
\left\{-p_{r,eq}+\frac{p_{w,eq}}{2k^w_+}
\left[q+\sqrt{q^2+4k^r_+ k^w_+}\right]\right\}e^{\lambda_+ t}
}{2k_r^+\left(4+\frac{q^2}{k^r_+ k^w_+}\right)}
\nonumber\\
&-&\frac{N_+\left(q-\sqrt{q^2+4k^r_+k^w_+}\right)
\left\{-p_{r,eq}+\frac{p_{w,eq}}{2k^w_+}
\left[q-\sqrt{q^2+4k^r_+ k^w_+}\right]\right\}e^{\lambda_- t}
}{2k_r^+\left(4+\frac{q^2}{k^r_+ k^w_+}\right)}\nonumber
\end{eqnarray}
where we defined the eigenvalues
\begin{equation}
\lambda_\pm=\frac{-\Sigma\pm\sqrt{\Sigma^2-4c}}{2}
\end{equation}
with $\Sigma=k^r_++k^r_-+k^w_++k^w_- $ and
$c=(k^r_++k^r_-)(k^w_++k^w_-)-k^r_+k^w_+$, and the quantities
\begin{eqnarray}
q&=&(k^r_++k^r_--k^w_+-k^w_-) \nonumber\\
\mathcal{N}_\pm&=&2+\frac{q}{2k^r_+k^w_+}\left(
q
\pm \sqrt{q^2+4k^r_+ k^w_+}
\right).
\end{eqnarray}

We now study the sign of the derivative of the error $\eta(t)$.
Clearly, its sign is the same as the sign of the derivative
of the function 
\begin{equation}
f(t)=\frac{p_w(t)}{p_r(t)}=\frac{p_{w,eq}+\delta p_w(t)}{p_{r,eq}+\delta p_r(t)}.
\end{equation}
The derivative of $f(t)$, before any simplification, reads:
\begin{align}
f'&=D^{-2}\left[ 2 k_w^+(k_w^- - k_r^-)(k_w^-k_r^+ + k_r^-(k_w^-+k_w^+)) 
\e^{-(3\Sigma + \sqrt{Q})t/2}  \right]\nonumber\\
&\Big{\{ } -2Q^2\e^{(\Sigma + \sqrt{Q})t/2}
+\e^{\Sigma t} [ {k^r_-}^2 + {k^w_-}^2 
+ k^r_- (-2 k^w_- + 2k^r_+ - 2 k^p_w + \sqrt{Q} ) \nonumber\\
&+ (k^r_+ + k^w_+) ( k^r_+ + k^w_++ \sqrt{Q}) 
+ k^w_- (-2 k^r_+ + 2k^w_+ +\sqrt{Q}) ]\nonumber\\
&+ \e^{(\Sigma +\sqrt{Q})t} [ {k^r_-}^2 + {k^w_-}^2 
- k^w_- (-2 k^w_+ + 2k^r_+ + \sqrt{Q} ) \nonumber\\
&- (k^r_+ + k^w_+) (- k^r_+ - k^w_++ \sqrt{Q}) + k^r_- 
(-2 k^r_+ + 2k^w_- + 2k^w_++\sqrt{Q}) ] \Big{\} }
\end{align}
where $\Sigma$ and $q$ are defined above, and we have also defined the
function of the rates $Q=q^2+4k^r_+ k^w_+$. The denominator $D$ has a
complicated expression that we omit since it is squared, hence it is
always positive. In the nominator, the term in square brackets clearly
has the sign of $k_-^w -k_-^r$. We now move to the study the sign of
the term in curly brackets, which can be expressed in terms of
hyperbolic functions as
\begin{equation}
\{\}= \e^{\frac12(\Sigma + \sqrt{Q})t}\left( -2 Q + 2\e^{\Sigma\frac t2} 
\left[Q\cosh(\sqrt{Q}t/2) - \sqrt{Q}\Sigma\sinh(\sqrt{Q}t/2)\right] \right)
\end{equation}
The prefactor is positive, and we are left with the two terms inside
the parenthesis. The first is clearly negative. To determine the sign
of the second, particularly the term in brackets, one should note that
$\sinh(x)>\cosh(x)$ for all positive $x$, and that $\Sigma>\sqrt{Q}$,
which is shown by expanding the squares. As a consequence, the second
term in the brackets is always larger than the first, so that the
whole term inside the parenthesis is negative. It follows that the
term in the curly brackets is negative, and that the sign of $f'$ is
just the sign of $k_-^r - k_-^w$. From this we conclude that the error grows
monotonically for all parameter choices.

Finally, the rates are parametrized in the main text by their kinetic and
energetic discriminations ($\delta$ and $\gamma$), the driving
$\epsilon$, and an overall time-scale $\omega$. The kinetic
discrimination $\delta$ appears in the forward rates, so that
$k_+^r/k_+^w=\e^{\delta}$. The driving $\epsilon$ is defined for right
bases, so that $k_+^r/k_-^r=\e^{\epsilon}$. Finally, the energetic
discrimination $\gamma$ reduces the driving of wrong bases, 
$k_+^w/k_-^w=\e^{\epsilon-\gamma}$. Summarizing, we have:
\begin{equation}\label{rates}
k_+^r=\omega \e^{\epsilon+\delta} \;\; ; \;\;k_-^r=\omega \e^\delta\;\; ; \;\; 
k_+^w=\omega \e^\epsilon \;\; ; \;\; k_-^w=\omega \e^\gamma .
\end{equation}

The condition $k_-^r - k_-^w$ is then equivalent to $\delta>\gamma$,
which we termed the case of {\em kinetic discrimination}. Conversely, when
$k_-^r < k_-^w$ (equivalent to$\delta<\gamma$) the error decreases
monotonically (regime of {\em energetic
discrimination}).

\section{Co-polymerization model}
\label{copol}
Our model is a minimal description of co-polymerization, similar to
that introduced in \cite{bennett79} and recently studied in
\cite{esposito} among others. As shown in the next section, it also
has strong parallelisms with Hopfield's original DNA copying scheme
\cite{hopfield74}. The setup consists of a template polymer chain
(such as a DNA strand), a growing copy of it (the newly formed strand
before cell division), and a chemically driven polymerase assisting
the process (as can be Pol$\gamma$). The polymerase adds and removes
monomers to the tip of the growing strand trying to match the monomer
sequence of the template strand.  Discrimination is performed in two
ways: rates of addition for correctly matching monomers are larger
than for incorrectly matching ones, and incorrect monomers are in a
high energy state (they are hence easier to remove). Finally, the
rates of addition of right/wrong monomers are higher than the
corresponding rates of removal, ensuring a net growth.

The rates for addition of a new monomer are expressed in
Eq. \ref{rates}. Energies $\delta$ and $\gamma$ allow for
discrimination through different barrier heights and different energy
of incorporation, respectively. The parameter $\epsilon$ corresponds
to the chemical driving. Finally $\omega$ determines the
time-scale. We denote a generic configuration of the chain by $[\&]$
(a state $[\&]$ can be thought of as a string of right and wrong
matches, for example $[\&]=[rwwwrwrrrw]$). At each time, three kinds
of events can occur: removal of the last element of the chain,
addition of a right base, or addition of a wrong base. Given the rates
in Eq. \ref{rates}, one can easily write the two rate equations for
this model as:
\begin{align}
\partial_t [\& r] & = [\&] k_+^r - [\& r] k^r_- \nonumber\\
\partial_t [\& w] &= [\&] k_+^w-[\& w] k^w_-
\end{align}
where states $[\& r]$ or $[\& w]$ are obtained from state $[\&]$ with
the addition of a right or wrong match respectively.  Following
Bennett's original approach \cite{bennett79}, we consider the steady
state in which there are constant fluxes of wrong $ \partial_t [\& w]=
J_w = \eta v [\&] $ and right $\partial_t [\& r]=J_r = (1-\eta) v [\&]
$ additions of aminoacids into the copied strain. Under these
assumptions, on can show \cite{bennett79,esposito} that the error is
given by:
\begin{equation}\label{err_cop}
\frac{k^w_+-k^w_- \eta}{k^r_+-k^r_-(1-\eta)}=\frac{\eta}{1-\eta}.
\end{equation}
while the average growth velocity of the copied strand is
\begin{equation}
v=k^w_+-k^w_-\eta+k^r_+-k^r_-(1-\eta).
\end{equation}
Using the same parametrization of the previous model,
Eq. (\ref{rates}), we write the chemical driving $\epsilon$ as a
function of the energetic discrimination energy $\gamma$, the kinetic
discrimination energy $\delta$, and the steady state error $\eta$. The
expression reads
\begin{equation}\label{eqeps}
\epsilon = \log\left[ \eta(1-\eta) \frac{1-e^{\gamma-\delta} }
{\eta - (1-\eta)e^{-\delta}} \right].
\end{equation}
By means of (\ref{eqeps}), the velocity can be expressed as
\begin{equation}\label{eqvel}
v=\omega \frac{1-(1+e^\gamma)\eta}{\eta-(1-\eta)e^{-\delta}}.
\end{equation}
We now want to impose that 1) the argument of the logarithm in
(\ref{eqeps}) has to be positive, and 2) the average velocity
(\ref{eqvel}) should also be positive. Assuming of course $0<\eta<1$,
the first condition is equivalent to
$(\delta-\gamma)[\eta-(1+e^\delta)^{-1}]>0$, while the second is
equivalent to
$[(1-e^\gamma)^{-1}-\eta][\eta-(1+e^\delta)^{-1}]>0$. Combining these
two conditions leads to Eq. (4) in the main text.

The entropy production rate $\dot{S}$ can be calculated with the usual
Schnakenberg formula \cite{schn}, that is
\begin{align}
\dot{S} = ([\&] k_+^r - [\& r] k^r_-) \log\left[\frac{[\&] k_+^r }{[\& r] k_r^-}\right]+ ([\&] k_+^w - [\& r] k^w_-)\log\left[\frac{[\&] k_+^w }{[\& r] k_-^w}\right]
\end{align}
 Using the expressions above, it is straightforward to show that the dissipation per step $\Delta S = \dot{S}/v$ is given by:
\begin{eqnarray}
\Delta S&=&\eta  \log\left[\frac 1\eta\right] + (1-\eta) \log
\left[\frac {1}{1-\eta}\right] +\eta  \log\left[\frac {k_{+} ^{w}}
{k_{-} ^{w}}\right] 
+ (1-\eta)  \log\left[\frac {k_{+} ^{r}}{k_{-} ^{r}}\right] \nonumber\\
&=& -\left(\eta \log\left[\eta\right] + (1-\eta) \log
\left[1-\eta\right]\right)   +(1-\eta)\epsilon+
\eta(\epsilon-\gamma)   ,
\end{eqnarray}
which can be expressed as a function of $\eta$, $\delta$ and $\gamma$
only by using Eq. (\ref{eqeps}).

Finally, notice that in the copolymerization model incorporation of
monomers and forth/back stepping of the polymerase are tightly coupled
together. However, this is not crucial for achieving the main results
of our paper, as we will discuss at the end of the last section of
this note.

\section{Mapping of Hopfield's original model} 
In Hopfield's formulation \cite{hopfield74}, given the template $c$,
by interacting through $C$ and $D$, either the aminoacid $P_C$ or
$P_D$ can be added to an RNA chain. Addition of $P_C$ will be the
right addition, and addition of $P_D$ will be considered an error. The
rate equation and steady state solution are:
\begin{equation}
c+C \xrightleftharpoons[k_C]{k'_C} [cC] \xrightharpoonup{v} P_C\;\;\;\;\; 
\mathrm{and} \;\;\;\;\; v[Cc] = k'_C [C] - k_C[Cc]
\end{equation}
and analogously for $D$. It is assumed that $[C] \sim [D]$, and
defined $f_C=[Cc]/[C]$ and $f_D=[Dc]/[D]$ as the fraction of
incorporated $C$ and $D$ monomers given a template $c$. At steady
state $f_C=1-f_D$, and $f_D$ is the error $\eta = f_D$. Solving the
system above we arrive at
\begin{equation}
\frac{\eta}{1-\eta} = \frac{f_D}{f_C} = \frac{k'_D}{k'_C}\frac{v + k_C}{v + k_D}
\end{equation}
Identifying these rates with those in our model according to Fig. 1,
the mapping to Hopfield's model is finished: $k_C = k_{-} ^{r}$, $k'_C
= k_{+} ^{r}$, $k_D = k_{-} ^{w}$ and $k'_D = k_{+} ^{w}$.

To verify the mapping we study two limiting cases. For $\gamma=0$ (as
Bennett assumed in \cite{bennett79}) we have that if $v\to\infty$,
then $\eta/(1-\eta)\to k'_D/k'_C = e^{-\delta}$; and if $v\to 0$, then
$\eta/(1-\eta)\to k'_Dk_C/k'_C k_D= 1$. On the other hand for
$\delta=0$ (as Hopfield assumed in \cite{hopfield74}) we have that if
$v\to\infty$, then $\eta/(1-\eta)\to k'_D/k'_C = 1$; and if $v\to 0$,
the classical result is obtained $\eta/(1-\eta)\to k'_Dk_C/k'_C k_D=
k_C/k_D=e^{-\gamma}$, in exact agreement with the
results obtained above.

\section{Proofreading model without intermediate state}

A minimal model of Kinetic Proofreading (KP) requires at least two
different pathways. The first is the copying pathway introduced above,
characterized by a driving which tends to make the chain grow. On the
other hand, the driving of the second pathway is {\em backward}, thus
undoing copies on average. The copying pathway has a bias towards
adding right bases by a faster (kinetic) and more stable (energetic)
binding. Conversely, the proofreading pathway has a bias towards
removing wrong bases by a faster and less stable unbinding. The
combination of both can reduce the minimal error given by the standard
copy, by the discrimination factor of the proofreading pathway. The
simplest proofreading scheme consists of the copying scheme introduced
before, and a parallel reaction which we characterize by four
additional proofreading rates $\tilde{k}_{\pm} ^{r/w}$.

\subsection{Rates parametrization}

\begin{figure}[htb]
\begin{center}
\includegraphics[width=10cm]{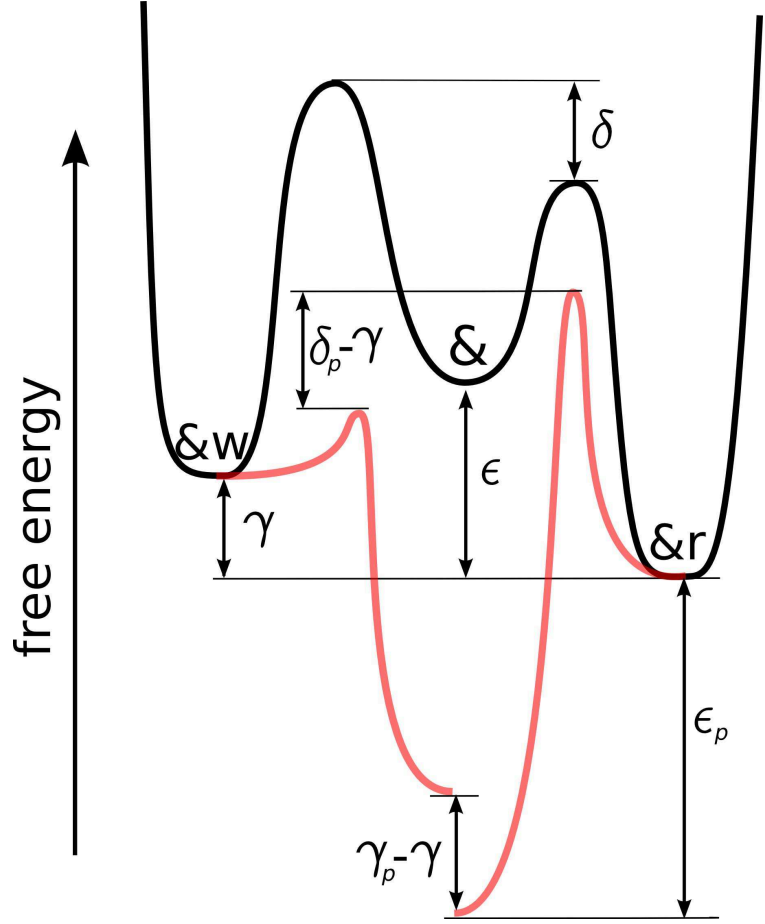}
\caption{Energy diagram of the reactions corresponding to the
  proofreading scheme with no intermediate steps, model C in the main
  text.\label{proof_benn}}
\end{center}
\end{figure}

We choose the same copying rates of the standard copying scheme, see
Eq.(\ref{rates}).  Further, we introduce proofreading rates which are
analogously characterized by a kinetic and energetic proofreading
discrimination factors ($\delta_p$ and $\gamma_p$), a backward driving
$\epsilon_p$, and an additional time-scale $\omega_p$.  In the case of
proofreading, we define the driving in the backward right additions,
that is $\tilde{k}^{r}_-/\tilde{k}^{r}_+ =\e^{\epsilon_p}$. The
kinetic discrimination is also backwards, and so
$\tilde{k}^{w}_-/\tilde{k}^{r}_- =\e^{\delta_p}$. Finally, the
energetic discrimination is reflected in a higher backward driving of
wrong bases, such that $\tilde{k}^{w}_-/ \tilde{k}^{w}_+
=\e^{\epsilon_p+\gamma_p}$. One can then write the proofreading rates
as
\begin{equation}\label{proofrates}
\tilde{k}^{r}_-=\omega_p \e^{\epsilon_p-\delta_p} \;\; ; \;\;
\tilde{k}^{r}_+=\omega_p \e^{-\delta_p}\;\; ; \;\; 
\tilde{k}^{w}_-=\omega_p \e^{\epsilon_p} \;\; ; \;\; 
\tilde{k}^{w}_+=\omega_p \e^{-\gamma_p}.
\end{equation}
The energy levels corresponding to this parametrization of the rates
are illustrated in Fig. \ref{proof_benn}. Notice that the end-states
in the proofreading pathway have a difference in energy
$\gamma-\gamma_p$. While in some coarse grained models such a
behaviour may be justifiable through external agents, typically one
would expect this difference not to exist, so that in the main text we
always fixed $\gamma_p=\gamma$. Further, we anticipate that numerical
results show that the proofreading step is always kinetic. This means
that the value of $\gamma_p$, as soon as it is positive, will not
anyway affect the minimum error achievable by the system.

\maketitle 

\subsection{Solving the model}
The kinetic equations in this case are:
\begin{align}
  \partial_t [\& r] & = [\&] ( k_+^r + \tilde{k}^{r}_+ ) - [\& r]
  (k^r_- +\tilde{k}^{r}_- )
  \nonumber\\
\partial_t [\& w] &= [\&] (k_+^w + \tilde{k}^{w}_+) - [\& w] 
(k^w_-+\tilde{k}^w_-).
\end{align}
Also in this case, the steady state solution can be obtained by
considering the fluxes of right and wrong bases added: $ \partial_t
[\& w]= J_w = \eta v [\&] $ and right $\partial_t [\& r]=J_r =
(1-\eta) v [\&] $. The error as a function of the rates is analogous
to the one for simple copying:
\begin{equation}\label{err_cop2}
\frac{k^w_++\tilde{k}^{w}_+- \eta(k^w_- + \tilde{k}^{w}_-)}
{ k^r_++\tilde{k}^r_+-(1-\eta)(k^r_-+\tilde{k}^r_-)}=\frac{\eta}{1-\eta}.
\end{equation}
The next step is to derive from this expression the driving
$\epsilon$ as a function of the error, the discriminations, and the
two new additional parameters: the proofreading driving $\epsilon_p$
and its characteristic time scale $\omega_p$. The result is:
\begin{align}
  \epsilon=\log\left[ \frac{1}{1-\eta(1+\e^\delta)}\left\{
      \eta(1-\eta)(\e^\gamma - \e^\delta + \omega_p\e^{\epsilon_p}
      -\omega_p\e^{\epsilon_p-\delta_p})-\omega_p
      e^{-\gamma_p}+\eta\omega_p(\e^{-\gamma_p}+e^{-\delta_p})
    \right\} \right]
\end{align}

The velocity is also analogous to that of the simple copying scheme:
\begin{equation}
v=k^w_+ + \tilde{k}^w_+-\eta(k^w_-+\tilde{k}^w_-)
+k^r_++\tilde{k}^r_+-(1-\eta)(k^r_- + \tilde{k}^r_-).
\end{equation}
However, for the entropy production rate, one has to consider the
transitions correponding to the two pathways independently:
\begin{eqnarray}
\dot{S} &=& (k_{+} ^w-\eta k_{-} ^w)\log\left[\frac{k_{+} ^{w}}{\eta k_{-} ^{w}}\right] + 
(k_+^{r}-(1-\eta)k_-^r)\log\left[\frac{k_+^r}{(1-\eta)k_-^r}\right]\nonumber\\
&+&(\tilde{k}^{w}_+-\eta \tilde{k}^w_-)\log\left[\frac{\tilde{k}^w_+}{\eta \tilde{k}^w_-}\right]
+ (\tilde{k}^r_+-(1-\eta)\tilde{k}^r_-)\log\left[\frac{\tilde{k}^r_+}{(1-\eta)\tilde{k}^r_-}
\right].
\end{eqnarray}
Finally, the dissipation per step is simply calculated as $\Delta S =
\dot{S}/v$.

\subsection{Minimization procedure and numerical results}

For each given value of the error $\eta$ and the four parameters
$\gamma$, $\gamma_p$, $\delta$, $\delta_p$, we identified the values
of the two remaining free parameters $\omega_p$ and $\epsilon_p$
corresponding to the minimum dissipation per step. In order to avoid
local minima, we adopted a sistematic minimization scheme: the two
parameters have been varied with a logarithmic step equal to $1.04$,
in an interval $10^{-5}<\omega_p,\epsilon_p<10^9$.  In this region,
we found the minimum dissipation per step with the constraint of a
positive reaction velocity. We also checked a posteriori that no
minimum was found at the boundaries of the minimization region.

\begin{figure}[htb]
\begin{center}
\includegraphics[width=14cm]{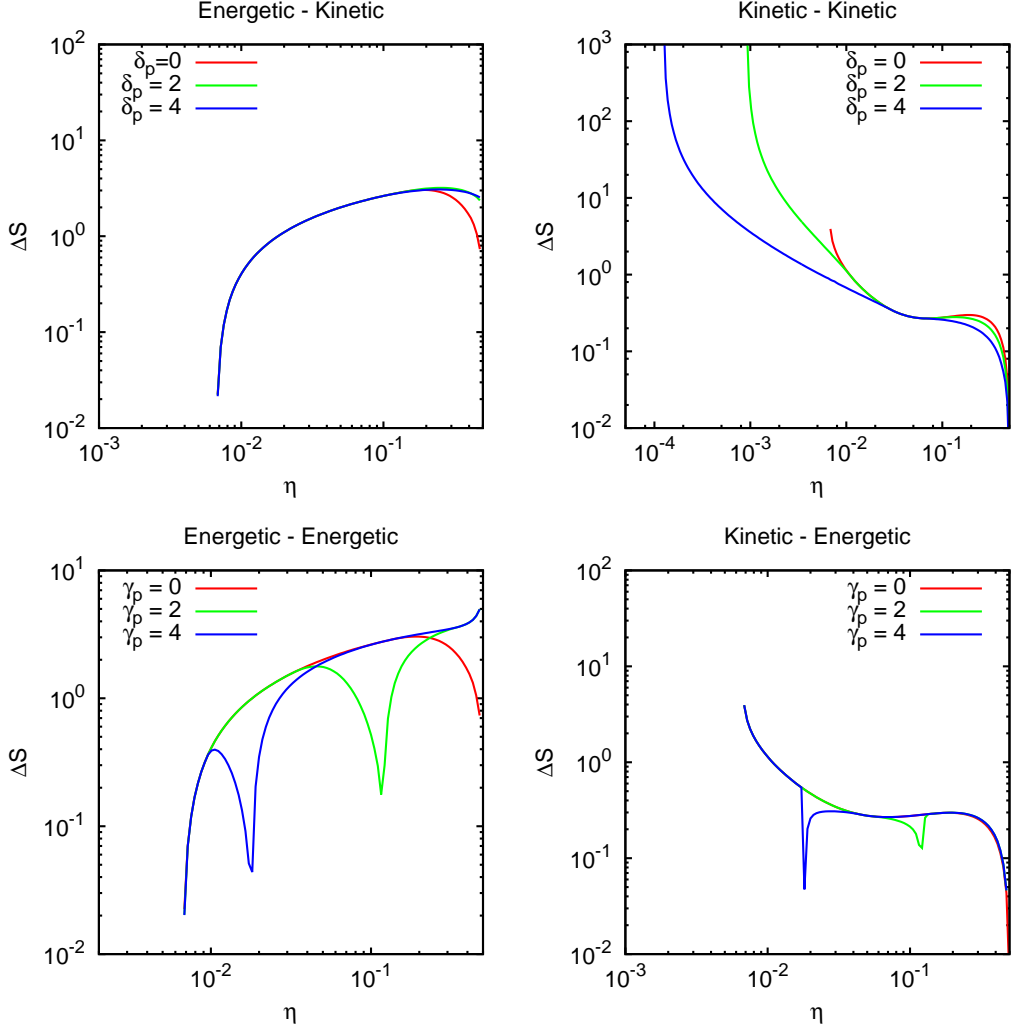}
\caption{Study of the four possible combination of energetic/kinetic
  discrimination and energetic/kinetic proofreading in Bennett's
  model. In the left panels the copy is energetic; in particular we
  chose $\delta=0$ and $\gamma=5$. Conversely, in the right panels the
  copy is kinetic with $\delta=5$ and $\gamma=0$. In the top panels the
  proofreading scheme is purely kinetic ($\gamma_p=0$), while in the
  bottom panel we fixed $\delta_p=0$ and varied $\gamma_p$. \label{fig_model3}}
\end{center}
\end{figure}

A sistematic simulation study of the $4$ possibilities of
energetic/kinetic copy coupled to energetic/kinetic proofreading is
presented in Fig. \ref{fig_model3}.  The results allows us for
reaching the following conclusions:

\begin{itemize}
\item The proofreading pathway can reduce the minimum error in the
  kinetic regime only. This can be seen in the lower panels of
  Fig. \ref{fig_model3}, where increasing $\gamma_p$ does not affect
  the minimum achievable error. In particular, in the bottom left
  panel the copying is energetic and the minimum error is given by
  $f(\gamma)=e^{-\gamma}/(1+e^{-\gamma})\approx 0.0067$ for
  $\gamma=5$. In the bottom right panel, the copying is kinetic and
  again the minimum error is given by $f(\delta)\approx 0.0067$ for
  $\delta=5$. The minima in the two figures correspond to parameters
  such as the proofreading reactions has an average forward flux
  instead of backward, so that the proofreading pathway works as an
  effective parallel adiabatic (energetic) copying pathway.

\item cooperative error reduction only takes place when both pathways
  are in the kinetic region. In the top right panel, increasing
  $\delta_p$ does not reduce the error. The only case in which the
  error can be reduced is in the kinetic-kinetic case of the top right
  panel, where the minimum error is given by
  $f(\delta)f(\delta_p)\approx f(\delta+\delta_p)\approx 0.0067,
  0.0009, 0.0001$ for $\delta_p=0, 2, 4$ respectively. We remark that
  this feature is a peculiarity of this model. We will show in the
  next section how including an intermediate state in the copying
  pathway allows for error reduction with an energetic copy and a
  kinetic proofreading.
\end{itemize}

\section{Proofreading model with intermediate state}
\label{kpinter}

In this section we present more extensive results on model $4$ of the
main text. This model presents some analogies with the previous one,
except that copying occurs via an intermediate state, denoted with a
``*'', which is connected with the final state in which the aminoacid
is incorporated.  This final state has also a proofreading step. The
overall reaction scheme is more in the spirit of Hopfield's original
proofreading mechanism.

\subsection{Parametrization of the rates}
The forward copying rates from the unbound to the intermediate state
are defined in exactly the same way as the copying rates in the
previous models, see Eq. (\ref{rates}). As in Hopfield's original
model, the transition rates from the intermediate state to the final
state have no discrimination, but have their own driving $\epsilon^*$
and time scale given by $\omega^*$.  They obey the relations
$\bar{k}^{w}_+/\bar{k}^{r}_+ =1$, $\bar{k}^{w}_+/\bar{k}^{w}_-= \e^{\epsilon^*}$ and
$\bar{k}^{r}_+/\bar{k}^{r}_-=\e^{\epsilon^*}$. It is not hard to show that
adding a discrimination below that of the original copying does not
reduce the error beyond the critical error. Adding a bigger one simply
reduces it to the critical error of this secondary copy, unlike the
additive effect of proofreading. The rates can be simply written as:
\begin{equation}\label{nondiscr}
\bar{k}^{r}_+ =\omega^* e^{\epsilon^*} \;\;\; ; \;\;
\bar{k}^{r}_- =\omega^*  \;\;\; ; \;\;\; 
\bar{k}^{w}_+ =\omega^* e^{\epsilon^*} \;\;\; ; \;\;\; 
\bar{k}^{w}_- =\omega^*    
\end{equation}

\begin{figure}[htb]
\begin{center}
\includegraphics[width=14cm]{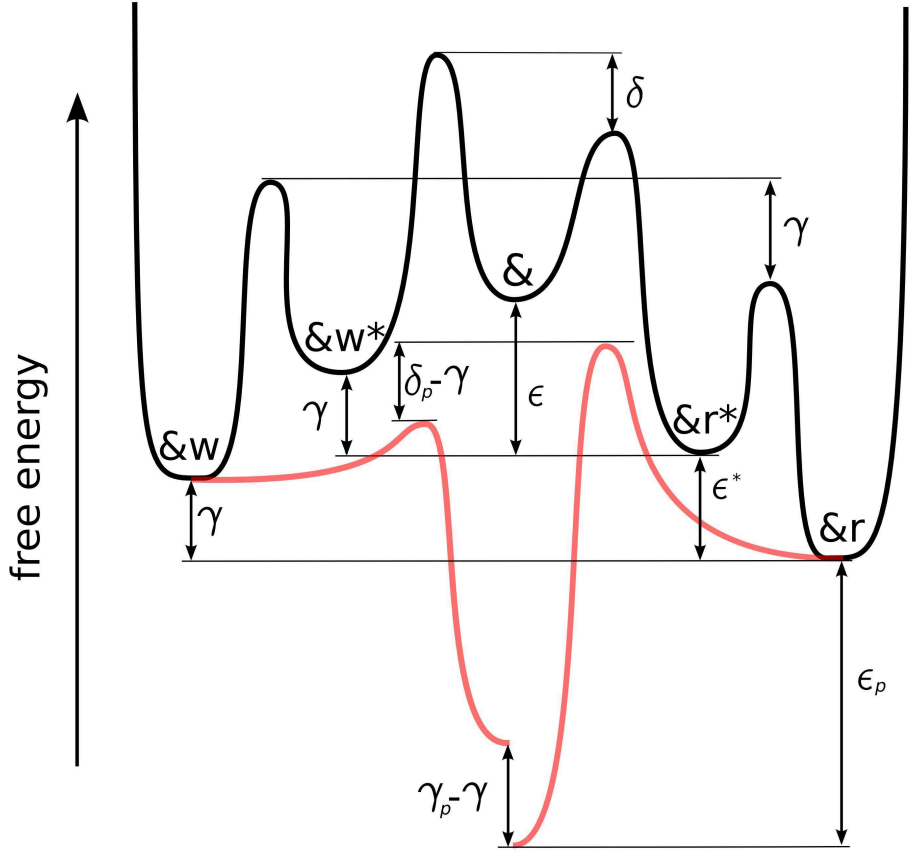}
\caption{Energy diagram of the reactions corresponding to the
  proofreading scheme with an intermediate step, model D in the main
  text.\label{mod4_scheme}}
\end{center}
\end{figure}

The final state is then connected with the initial state by the same
proofreading rates defined in the previous section,
Eq. (\ref{proofrates}).  The full energy diagram is depicted in
Fig. \ref{mod4_scheme}. As before, the energy difference
$\gamma-\gamma_p$ is irrelevant as the proofreading step has to be a
kinetic step, and so we choose it arbitrarily to be null. Again, this
corresponds to the physical requirement that the energy of the chain
can not change if no base is added.

\title{Proofreading a la Hopfield}
\maketitle 

\subsection{Solving the model}
With the notation introduced in the previous section, it is easy to
write the four kinetic equations of this proofreading scheme:
\begin{eqnarray}
\partial_t [R] &=& [R^*] \bar{k}^{r}_+ + [\&]\tilde{k}^{r}_+ - [R] ( \bar{k}^{r}_- +\tilde{k}^{r}_-)\nonumber\\   
\partial_t[W] &=& [W^*] \bar{k}^{w}_+ + [\&]\tilde{k}^{w}_+ - [W] ( \bar{k}^{w}_- +\tilde{k}^{w}_-)\nonumber\\
\partial_t[R^*] &=& [\&] k^r_+ + [R] \bar{k}^{r}_- - [R^*] ( k^r_- + \bar{k}^{r}_+)\nonumber\\
\partial_t[W^*] &=& [\&] k^w_- + [W] k^{w*}_- - [W^*] ( k^w_- + \bar{k}^{w}_+).
\end{eqnarray}
The easiest way to obtain the solution is by flux balance at the
steady state of constant growth velocity $v$, which corresponds to:
\begin{align}
[W] v &=([\&] \tilde{k}^{w}_+ - [W]\tilde{k}^{w}_-) + ([W^*]\bar{k}^{w}_+ - [W] \bar{k}^{w}_- ) \nonumber\\
[\&] k^w_+ - [W^*] k^w_-&= [W^*]\bar{k}^{w}_+ - [W] \bar{k}^{w}_- \nonumber\\
[R] v &=([\&] \tilde{k}^{r}_+ - [R]\tilde{k}^{r}_-) + ([R^*]\bar{k}^{r}_+ - [R] \bar{k}^{r}_-) \nonumber\\
[\&] k^r_+ - [R^*] k^r_-&= [R^*]\bar{k}^{r}_+ - [R] \bar{k}^{r}_- .
\end{align}
As before, we seek equations to determine the error rate and the
velocity as a function of the rates.  We proceed by dividing each of
the equations in section by $\&$ and define $W/\&=\eta$,
$R/\&=(1-\eta)$, $W^*/\&=w^*$, $R^*/\&=r^*$. By means of the 2nd and
4th equations we find an expression for $r^*$ and $w^*$:
\begin{align}
w^*=\frac{k^w_++\eta \bar{k}^{w}_-}{k^w_-+\bar{k}^{w}_+}\nonumber\\
r^*=\frac{k^r_++(1-\eta) \bar{k}^{r}_-}{k^r_-+\bar{k}^{r}_+}.
\end{align}
Substituting into the other 2 equations lead to two coupled equations
for $\eta$ and $v$.
\begin{eqnarray}
\label{ecuaciones}
  \eta v&=&(\tilde{k}^{w}_+-\eta \tilde{k}^{w}_-)
+\left[\bar{k}^{w}_+\frac{k^w_++\eta \bar{k}^{w}_-}{k^w_-+\bar{k}^{w}_+}
-\eta \bar{k}^{w}_-\right]\nonumber\\
(1-\eta)v&=&[\tilde{k}^{r}_+-(1-\eta) \tilde{k}^{r}_-]
+\left[\bar{k}^{r}_+\frac{k^r_++(1-\eta) \bar{k}^r_-}{k^r_-+\bar{k}^{r}_+}
-(1-\eta) \bar{k}^{r}_-\right].
\end{eqnarray}
Now we multiply the first equation by $(1-\eta)$, the second by $\eta$
and subtract the second from the first to find a closed expression for
$\eta$:
\begin{eqnarray}
(1-\eta)(\tilde{k}^{w}_+-\eta \tilde{k}^{w}_-)+(1-\eta)
\left[\bar{k}^{w}_+\frac{k^w_++\eta \bar{k}^{w}_-}{k^w_-+\bar{k}^{w}_+}-\eta \bar{k}^{w}_-\right]\nonumber\\
-\eta [\tilde{k}^{r}_+-(1-\eta) \tilde{k}^{r}_-]
-\eta \left[\bar{k}^{r}_+\frac{k^r_++(1-\eta) \bar{k}^{r}_-}{k^r_-+\bar{k}^{r}_+}
-(1-\eta) \bar{k}^{r}_-\right]=0.
\end{eqnarray}

Again, this formula can be inverted to obtain the copying driving $\epsilon$ as a
function of $\eta$ and the other energy differences:
\begin{equation}
\e^\epsilon=  \frac{\eta\omega_p\e^{-\delta_p} - (1-\eta)\omega_p\e^{-\gamma_p}  
+\eta(1-\eta)\left(  \omega_p\e^{\epsilon_p}(1-\e^{-\delta_p}) + 
\frac{(\omega^{*2} \e^{\epsilon^*})(\e^\gamma-\e^\delta)}{(\e^\delta+\omega^*\e^{\epsilon^*})
(\e^\gamma+\omega^*\e^{\epsilon^*})}
\right)}{\omega^*\e^{\epsilon^*}\left( \frac{1-\eta}{\e^\gamma+
\omega^*\e^{\epsilon^*}}  - \frac{\eta\e^\delta}
{\e^\delta+\omega^*\e^{\epsilon^*}} \right)}.
\end{equation}

The velocity is straightforward to calculate from one of the
expressions in (\ref{ecuaciones}), and is simply:
\begin{equation}
v=\left(\frac{\tilde{k}^w_+}{\eta}- \tilde{k}^w_-\right)
+\left[\frac{\bar{k}^w_+}{\eta}\frac{k^w_++\eta \bar{k}^{w}_-}{k^w_-+\bar{k}^{w}_+}- \bar{k}^{w}_-\right]
\end{equation}

Finally, we calculate the entropy production by summing the six
contributions of the local fluxes of the system. This results in the
following lengthy expression:
\begin{eqnarray}
 \& \dot{S} &=&(\& k^w_+ - W^* k^w_-) \log\left[\frac{\& k^w_+ }{W^* k^w_-}\right] 
+ (W^* \bar{k}^{w}_+ - W \bar{k}^{w}_-) \log\left[\frac{W^* \bar{k}^{w}_+ }{W \bar{k}^{w}_-}\right] \nonumber\\
&+& (\& \tilde{k}^{w}_+ - W \tilde{k}^{w}_-) \log\left[\frac{\& \tilde{k}^{w}_+ }{W \tilde{k}^{w}_-}\right]
  + (\& k^r_+ - R^* k^r_-) \log\left[\frac{\& k^r_+ }{R^* k^r_-}\right] \nonumber\\
&+& (R^* \bar{k}^{r}_+ - R \bar{k}^{r}_-) \log\left[\frac{R^* \bar{k}^{r}_+ }{R \bar{k}^{r}_-}\right] 
+ (\& \tilde{k}^{r}_+ - R \tilde{k}^{r}_-) \log\left[\frac{\& \tilde{k}^{r}_+ }{R \tilde{k}^{r}_-}\right] .
\end{eqnarray}
Dividing by $\&$ and using the expressions for $r^*$, $w^*$ and $\eta$, 
we obtain the rate of entropy production:
\begin{eqnarray}
\dot{S} &=& 
( k^w_+ - \frac{k^w_++\eta \bar{k}^{w}_-}{k^w_-+\bar{k}^{w}_+}k^w_-) \log\left[\frac{ (k^w_-+\bar{k}^{w}_+) k^w_+ }{(k^w_++\eta \bar{k}^{w}_-) k^w_-}\right] \nonumber\\
&+& (\frac{k^w_++\eta \bar{k}^{w}_-}{k^w_-+\bar{k}^{w}_+} \bar{k}^{w}_+ - \eta \bar{k}^{w}_-) \log\left[\frac{(k^w_++\eta \bar{k}^{w}_-) \bar{k}^{w}_+ }{(k^w_-+\bar{k}^{w}_+)\eta \bar{k}^{w}_-}\right]\nonumber\\
&+& ( \tilde{k}^{w}_+ - \eta \tilde{k}^{w}_-) \log\left[\frac{ \tilde{k}^{w}_+ }{\eta \tilde{k}^{w}_-}\right] \nonumber\\
&+&  ( k^r_+ - \frac{k^r_++(1-\eta) \bar{k}^{r}_-}{k^r_-+\bar{k}^{r}_+} k^r_-) \log\left[\frac{ (k^r_-+\bar{k}^{r}_+)k^r_+ }{(k^r_++(1-\eta) \bar{k}^{r}_-) k^r_-}\right]\nonumber\\
&+& (\frac{k^r_++(1-\eta) \bar{k}^{r}_-}{k^r_-+\bar{k}^{r}_+} \bar{k}^{r}_+ - (1-\eta) \bar{k}^{r}_-) \log\left[\frac{(k^r_++(1-\eta) \bar{k}^{r}_-)\bar{k}^{r}_+ }{(k^r_-+\bar{k}^{r}_+)(1-\eta) \bar{k}^{r}_-}\right] \nonumber\\
&+& ( \tilde{k}^{r}_+ - (1-\eta) \tilde{k}^{r}_-) \log\left[\frac{ \tilde{k}^{r}_+ }{(1-\eta) \tilde{k}^{r}_-}\right] .
\end{eqnarray}

\subsection{Minimization procedure and numerical results}

In analogy with the previous model, for each value of the parameters
$\delta$, $\delta_p$, $\gamma$ and $\gamma_p$ and the variable $\eta$
we found the values of the free parameters corresponding to the
minimum dissipation per step. In this case we had to minimize with
respect to four free parameters: $\omega_p$, $\epsilon_p$, $\omega^*$
and $\epsilon^*$. Given the number of parameters, we implemented a
larger logarithmic minimization step, equal to $1.2$.

\begin{figure}[htb]
\begin{center}
\includegraphics[width=14cm]{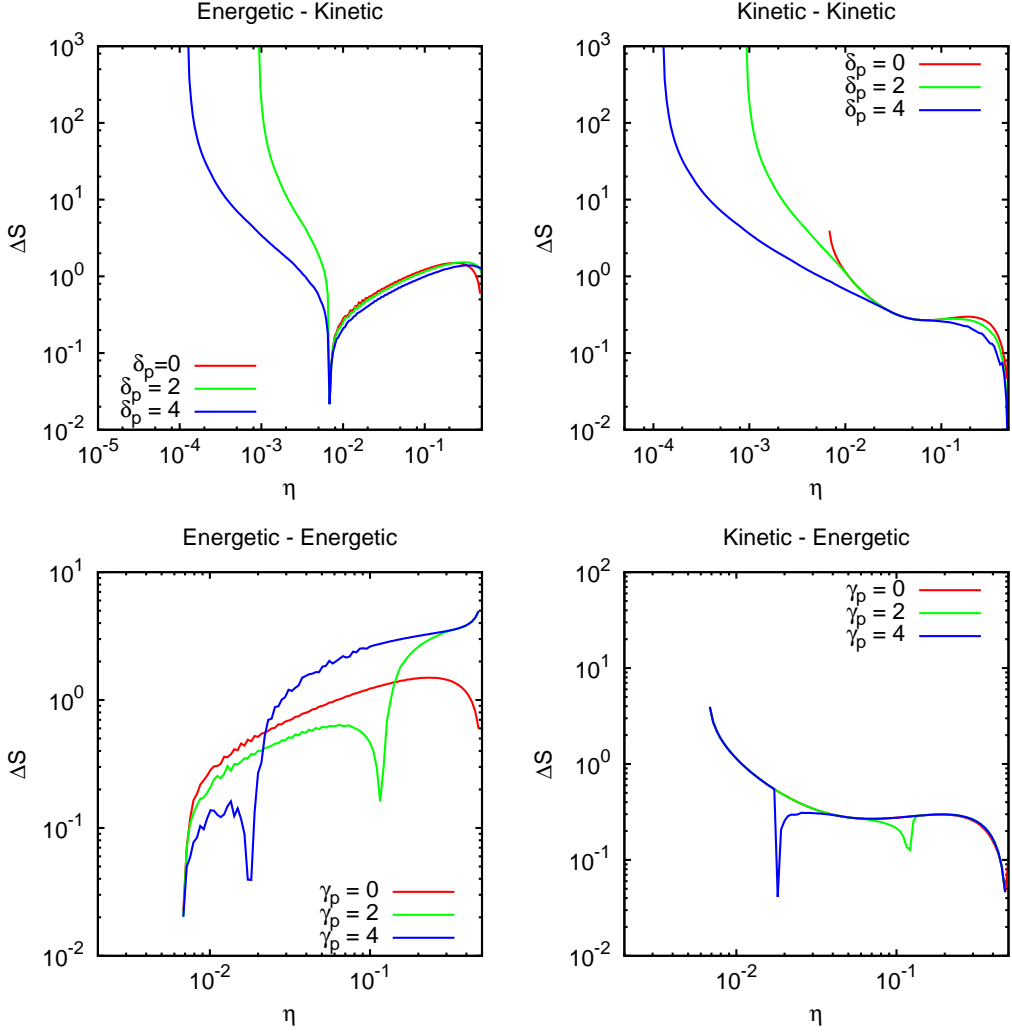}
\caption{Study of the four possible combination in the proofreading model 
with an intermediate state. In the left panels the copy is energetic; in particular we
  chose $\delta=0$ and $\gamma=5$. Conversely, in the right panels the
  copy is kinetic with $\delta=5$ and $\gamma=0$. In the top panels the
  proofreading scheme is purely kinetic ($\gamma_p=0$), while in the
  bottom panel we fixed $\delta_p=0$ and varied $\gamma_p$. 
  \label{fig_model4}}
\end{center}
\end{figure}

The result of Fig. \ref{fig_model4} are consistent to those of the
previous model, see Fig. \ref{fig_model3}. The only important difference is:

\begin{itemize}
\item The presence of an additional step in the copying pathway allows
  for error reduction via an energetic copy - kinetic proofreading
  scheme. This can be seen in the top left panel of
  Fig. \ref{fig_model4}, where the minimum error does depend on
  $\delta_p$ via the usual function $f(\gamma)f(\delta_p)\approx
  f(\gamma+\delta_p)$. This is at variance with model 3, shown in
  Fig. \ref{fig_model3}, where the minumum error in the same case
  was simply equal to $f(\gamma)$.
\end{itemize}

Finally, notice that this same model, but without the proofreading
pathway (i.e. with $\omega_p=0$) becomes a variant of the
copolymerization copying model with an intermediate step. The two
steps can be thus interpreted as the (discriminating) copying step,
characterized by the same rates as the copolymerization model, and a
moving, non-discriminating step, characterized by the rates in
Eqs. (\ref{nondiscr}). This model can thus be used to investigate
whether our results on the copolymerization model depend crucially on
the fact that movement and monomer incorporation are tightly linked
together by relaxing this assumption. The curves for
$\gamma_p=\delta_p=0$ in Fig. \ref{fig_model4} already suggest that
the minimum and maximum error in this limit should be still given by
Eq. (4) in the main text. Additional simulations (not shown) performed
with the constraint $\omega_p=0$ confirm this scenario. We can thus
conclude that the main results of the paper about the copolymerization
model are robust and independent of the simplifying assumption of
linking monomer incorporation and polymerase movement.

\section{Co-polymerization with decoupled stepping-copying}

\begin{figure}[htb]
\begin{center}
\includegraphics[width=14cm]{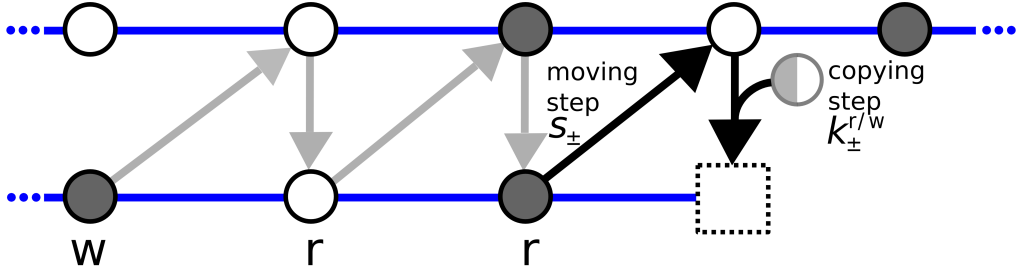}
\caption{Scheme of a polymerase copying a base in a model where
  stepping and copying are independent. The stepping rates are
  $s_{\pm}$, the copying rates are $k_\pm^{r/w}$.\label{walk} }
\end{center}
\end{figure}

In the co-polymerization model described in section \ref{copol}, it is
assumed that the polymerase moves forward/backward each time a base is
added/removed. In a more realistic model, the moving step and the
copying step are successive but independent one from the other. That
is, the polymerase moves to a new base with a mechanical step, copies
it through a chemical step, and goes on to the next base, as
represented in Fig. \ref{walk}. In this section, we study such variant
model to demonstrate that the assumption of coupled stepping/copying
steps made in the main text is not crucial for the results of our work.
 
\begin{figure}[htb]
\begin{center}
\includegraphics[width=14cm]{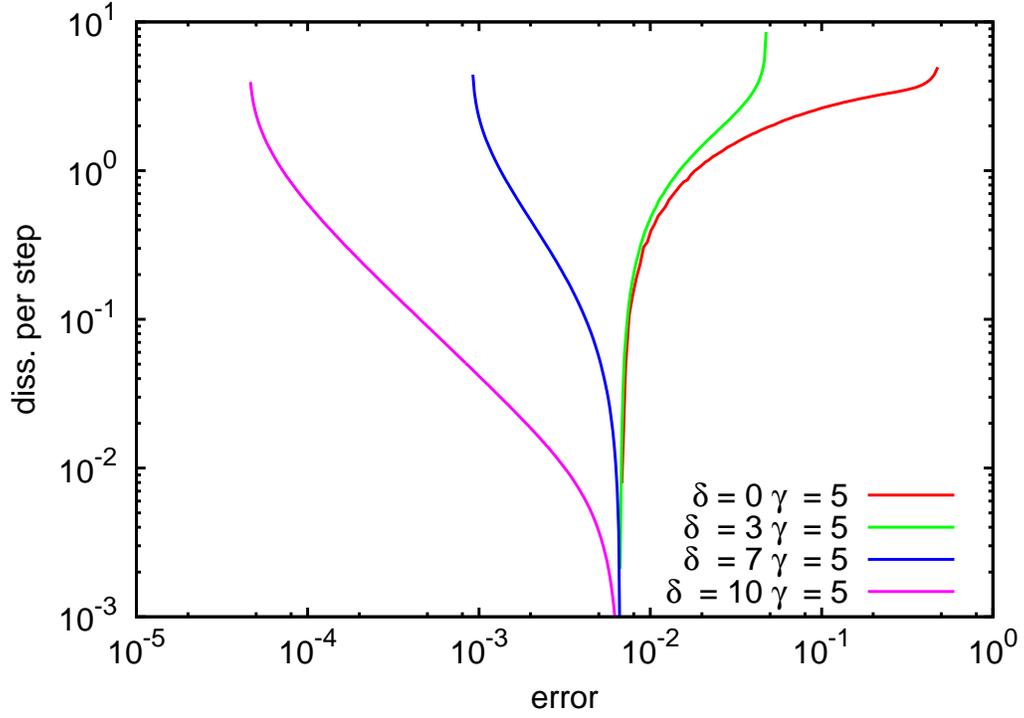}
\caption{Minimum dissipation per added base $\Delta S$ (combination of copying and moving steps) as a function of the error $\eta$, at fixed discrimination energy $\gamma$ varying the discriminating barrier $\delta$. As in the co-polymerization model, for $\delta<\gamma$ the minimum error occurs at $\e^{-\gamma}/(1 + \e^{-\gamma}) \approx 0.0067$ for $\gamma = 5$ (see Eq. 3 in the main text). This corresponds to the energetic region. For $\delta>\gamma$, the minimum error is $\e^{-\delta}/(1 + \e^{-\delta})$, approximately equal to 0.0009 for $\delta = 7$ and 0.000045 for $\delta = 10$.
\label{walkdiss}
}
\end{center}
\end{figure}

To describe a co-polymerization model with copying decoupled from
stepping, we need to specify stepping and copying rates. The copying
rates $k_{\pm}^{r/w}$ are simply given by Eq. \ref{rates} as
before. We assume the stepping rates to be independent of the binding
of a right/wrong monomer, and we parametrize them in the same way as
the rates in Eq. \ref{nondiscr}:
\begin{eqnarray}
s^+&=&\omega^* e^{\epsilon^*}\nonumber\\
s^-&=&\omega^*.
\end{eqnarray}
With this choice, stepping is a non-discriminatory process with a
chemical driving $\epsilon^*$ and a characteristic stepping time of
$1/\omega^*$. This model can be solved similarly to the proofreading
model in section \ref{kpinter} (it can actually be thought as a
particular case of the model in section \ref{kpinter} in the absence
of proofreading). A minimization procedure on the stepping parameters
$\{\epsilon^*, \omega^*\}$ analogous to that used in section
\ref{kpinter}, yields the results in Fig. \ref{walkdiss}.

It is clear that Fig. \ref{walkdiss} presents the same features
Fig. 3b of the main text, which corresponds to the co-polymerization
with tight coupling between stepping and copying. For values
$\delta<\gamma$ the system is in the energetic regime, and the minimum
error is $f(\gamma)$, while the maximum error is
$f(\delta)$. Conversely, for $\delta>\gamma$, in the kinetic regime,
the minimum error is given by $f(\delta)$ and the maximum error is
$f(\gamma)$. In other words, simulations show that the prediction of
Eq. 3 in the main text on the value of the minimum and maximum error
are still valid for this variant of the model: as stepping is a
non-discriminatory process, it does not affect the critical
errors. Furthermore, the trends of $\Delta S$ are preserved: in the
kinetic regime, the system becomes very dissipative upon approaching
the minimum error. In the energetic regime, the minimal error is
achieved at near-equilibrium conditions, where copying and stepping
are both slow.


\begin{thebibliography}{10}
\bibitem{supplements} Details on mathematical demonstrations and
    numerical simulations are available as an online appendix.
\end{thebibliography}
\end{document}